\documentclass{WileyMSP-template}
\usepackage{ragged2e}
\usepackage{amsmath}
\usepackage{xcolor}
\usepackage{soul}
\usepackage{siunitx}
\usepackage[dvipsnames]{xcolor}
\begin{document}
%\SetWatermarkText{DRAFT}
%\SetWatermarkScale{4}
%\SetWatermarkColor[gray]{0.90}

%\pagestyle{fancy}
%\rhead{\includegraphics[width=2.5cm]{vch-logo.png}}

\title{Scalable Cyclic Olefin Copolymer Encapsulation for High Optical Quality of TMD Monolayers}

\maketitle

% Author: Please give full first and last names for authors and include * after the name of all corresponding authors

\author{Suprova Das$^{1,2,3,4}$, Md Tarik Hossain$^{5,*}$, Zlata Fedorova$^{1,2,4,*}$, Zifei Zhang$^{1,2,4}$, Axel Printschler$^{5}$, \\ Begimai Adilbekova$^{5}$, Honey Jayeshkumer Shah$^{5}$, Stefan Velja$^{7}$,
Caterina Cocchi$^{2,7}$
Andrey Turchanin$^{2,5,6,*}$, Isabelle Staude$^{1,2,3,4}$}

\begin{affiliations}
\author{$^{1}$Institute of Solid State Physics, Friedrich Schiller University Jena, 07743 Jena, Germany\\
$^{2}$Abbe Center of Photonics, Friedrich Schiller University Jena, 07745 Jena, Germany\\
$^{3}$Max Planck School of Photonics, 07745 Jena, Germany\\
$^{4}$Institute of Applied Physics, Friedrich Schiller University Jena, 07745 Jena, Germany\\
$^{5}$Institute of Physical Chemistry, Friedrich Schiller University Jena, 07743 Jena, Germany\\
$^{6}$Jena Center for Soft Matter (JCSM), 07743 Jena, Germany\\
$^{7}$Institute of Condensed Matter Theory and Optics, Friedrich Schiller University Jena, 07743 Jena, Germany\\
$^{*}$Corresponding authors: tarik.hossain@uni-jena.de, zlata.fedorova@uni-jena.de, andrey.turchanin@uni-jena.de}
\end{affiliations}

\\

\begin{abstract}
\justifying
\noindent
Monolayer transition metal dichalcogenides (TMDs) combine a direct bandgap, strongly bound excitons, and pronounced second-order optical nonlinearity, which makes them promising materials for ultrathin optoelectronic and nanophotonic devices. However, their optical performance is often degraded by environmental exposure and substrate-induced charge trapping, motivating the development of scalable encapsulation strategies. Here, we investigate spin-coated cyclic olefin copolymer (COC) as a scalable encapsulant for TMDs. Room-temperature and cryogenic optical spectroscopy reveal enhanced photoluminescence and second-harmonic generation, accompanied by excitonic linewidth narrowing and an increased exciton-to-trion ratio. In addition, COC encapsulation induces an excitonic peak splitting and an overall spectral blueshift. First-principles calculations attribute these spectral modifications to local symmetry breaking at the chalcogen interface and macroscopic compressive strain, respectively. These findings establish spin-coated COC as an effective, scalable encapsulation strategy and a potential platform for post-growth excitonic and band-structure engineering.

\end{abstract}

\section{Introduction}

\justifying
Atomically thin transition-metal dichalcogenides (TMDs) monolayers (MLs) have garnered considerable attention owing to their exceptional optical and electronic properties~\cite{mak2016photonics,de2025roadmap}. In particular, MLs such as $\mathrm{Mo}\mathrm{S}_2$, exhibit a direct bandgap, giving rise to strong photoluminescence (PL) in the visible spectral range~\cite{splendiani2010emerging,mak2010atomically,mak2012control}. In addition, the broken inversion symmetry in TMD MLs enables intense second-harmonic generation (SHG) \cite{Malard2013,Kumar2013}. Strongly bound excitons in these materials produce sharp resonances that persist even at room temperature (RT) and as a result, excitonic effects govern many of their optical properties, including PL, absorption, reflection, and SHG. The optical response of TMD MLs is highly sensitive not only to their intrinsic quality \cite{Edelberg2019, Amani2015} but also to the surrounding environment. For example, the PL is often quenched due to unintended doping induced by the substrate. This effect is strongly influenced by the quality of the ML-substrate interface, including the density of trap states, substrate roughness, strain, and the local dielectric environment \cite{Ajayi2017,Shree2020,Ngo2025, Wang2018}. Furthermore, environmental effects \cite{Chandran2026,Gao2016} such as oxidation \cite{Gao2016,Kotsakidis2019} and photodegradation \cite{Chang2021} can further degrade the intrinsic quality of MLs. To address these issues, encapsulation strategies have been employed to reveal the intrinsic optical  performance of the TMD MLs.\\
Hexagonal boron nitride (hBN) is the most commonly used encapsulating material, as it provides an atomically flat and chemically inert interface that effectively protects the ML~\cite{Man2016,Ahn2016}. As a result, hBN encapsulation significantly enhances the optical quality  of 2D TMDs~\cite{Cadiz2017, Ye2018}: it sharpens excitonic resonances, and enables the observation of  fine excitonic features and many-body excitonic phenomena~\cite{Cadiz2017,Ye2018,Li2018,Li2019,Qian2024}. However, the scalability of hBN-based encapsulation remains limited, since high-quality hBN is predominantly obtained via mechanical exfoliation of ultrapure bulk crystals~\cite{Taniguchi2007}. In addition, the assembly of hBN/TMD heterostructures requires a complex transfer and stacking procedure, which is labor-intensive and challenging to scale.\\
Several other encapsulation strategies have been proposed to address the scalability limitations of hBN. For example, Ga$_2$O$_3$ provides large-area passivation and enhanced PL, but still relies on transfer-based processing~\cite{wurdack2021ultrathin}. Alternatively, wafer-scale oxide encapsulation by atomic layer deposition or physical vapor deposition is compatible with scalable device fabrication, although it often alters or degrades the excitonic properties of TMD MLs~\cite{ngo2025scalable}.
In this context, polymer-based approaches offer a promising alternative. Specifically, cyclic olefin copolymer (COC) has been shown to passivate the $\mathrm{Si}\mathrm{O}_2$ dielectric, resulting in significant enhancement of PL peak intensity in ML $\mathrm{Mo}\mathrm{S}_2$ at room temperature~\cite{kalkan2023high}. This establishes COC as an effective platform for modifying the dielectric environment and mitigating substrate-induced trap states.\\
In this work, we systematically investigate the impact of COC encapsulation on the linear and nonlinear optical response of ML $\mathrm{MoS}_2$. The COC layer is deposited by a simple spin-coating process that is fast, low-cost, and compatible with large-area fabrication. We demonstrate that COC encapsulation significantly enhances PL at both room and cryogenic temperatures, narrows excitonic linewidths, increases the exciton-to-trion ratio, and boosts the SHG response. In addition, COC encapsulation induces excitonic peak splitting and an overall spectral blueshift.
To elucidate the microscopic origin of these spectral modifications, we complement our experiments with density-functional theory (DFT) calculations, which identify local symmetry breaking at the interface and macroscopic compressive strain as the driving mechanisms behind the excitonic splitting and blueshift, respectively. Overall, these results establish COC encapsulation as a straightforward and scalable strategy for integrating high-performance TMD MLs into advanced photonic and optoelectronic platforms. Moreover, the observed encapsulation-induced spectral modifications suggest that COC may provide a new route toward post-growth engineering of the excitonic and electronic structure of TMDs.

\section{Results and Discussions}
\subsection{Room-temperature PL and Raman spectroscopy}
\justifying 
We studied the optical properties of two configurations: ML $\text{MoS}_2$ on glass and COC-encapsulated  $\text{MoS}_2$ ML on glass, as illustrated in the left and right panels of Figures 1a, respectively. Interactions with the surrounding environment, including oxygen, moisture, and substrate-induced trap states, can introduce non-radiative recombination channels and thereby reduce the optical quality of TMD MLs. To address this issue, we encapsulated the TMD MLs using ultrathin COC layers. To this end, ML $\text{MoS}_2$ was first synthesized on 300 nm $\text{SiO}_2$/Si substrates via a one-step chemical vapour deposition (CVD) method at atmospheric pressure employing solid phase $\text{MoO}_3$ and sulphur precursors as Mo and S sources, respectively~\cite{George2019}. The CVD-grown MLs were subsequently transferred onto bare and ultrathin COC-coated glass substrates using a PMMA-assisted transfer technique~\cite{Winter2018}. 
Atomic force microscopy (AFM) confirmed a COC thickness of $\sim$10~nm and revealed that the root-mean-square surface roughness of the  COC layer  is significantly lower compared to a bare glass substrate ($0.31 \pm 0.01$ nm vs $0.48 \pm 0.03$ nm), indicating that the COC layer provides an improved dielectric interface (see Figure S2 of the Supplementary Information (SI)). Further, a top layer of COC is deposited by spin-coating using the same procedure, resulting in a symmetric COC/TMD/COC heterostructure (see Methods section for details). Optical microscopy images of both configurations are shown in Figure S4, SI.

Next, we performed room-temperature Raman and PL spectroscopy and white light reflectivity to assess the effect of COC encapsulation on the optical quality of the MLs. Figure 1b presents the Raman spectra of $\text{MoS}_2$ on bare glass and COC-encapsulated $\text{MoS}_2$, displaying the characteristic $E'$ and $A'_1$ vibrational modes of ML $\text{MoS}_2$~\cite{lee2010anomalous}. After COC encapsulation, both the $E'$ and $A'_1$ modes show a noticeable blueshift, indicative of compressive strain in the ML~\cite{rice2013raman}. A comparison of the Raman peak positions of pristine and COC-encapsulated $\text{MoS}_2$ ML on glass  substrates is provided in Table S1 of the SI. To further assess the uniformity of the MLs, Raman intensity maps were recorded for the $A'_1$ mode of $\text{MoS}_2$ across the flake in both configurations (Figures 1c). The $A'_1$ peak intensity map of $\text{MoS}_2$ (Figure 1c, left) shows a uniform signal in  $\text{MoS}_2$ ML without COC encapsulation. In contrast, the sample with COC encapsulation shows less spatial uniformity, but an overall stronger signal of the $A'_1$ peak intensity map (Figure 1c, right). PL measurements further reveal a remarkable influence of the COC. With COC, the PL intensity increases by at least two times (Figure 1d), accompanied by a blue shift of the neutral A-exciton emission from 1.849 to 1.894 eV, as determined from pseudo-Voigt fits (Figure S6, SI). The blue shift is also observed in white-light reflectivity, where the A-exciton resonance moves from \SI{1.873}{\electronvolt} to \SI{1.891}{\electronvolt} (Figure S5, SI). The smaller shift in reflectivity together with the spectral broadening suggest that the room-temperature absorption resonance can contain unresolved overlapping transitions, which become distinguishable only in low-temperature measurements discussed later. Overall, the blue shift of A-exciton is consistent with compressive strain observed in Raman spectra, although additional contributions from the reduced substrate-induced doping may also be present.

 Figure 1e presents the PL intensity maps of ML $\text{MoS}_2$ on the bare substrate and the encapsulated sample, respectively, demonstrating an overall enhancement in PL intensity across the entire flake after encapsulation. This enhancement is attributed to suppressed charge transfer processes and a reduction in interface trap states~\cite{Kalkan2023}. Furthermore, the PL map of the encapsulated sample exhibits localized emission hotspots (Figure 1e right panel), which are likely associated with defect-bound excitons, as discussed below in relation to the low-temperature PL measurements in Figure 3. 

\begin{figure}[h]
  \centering
  \includegraphics[width=0.7\linewidth]{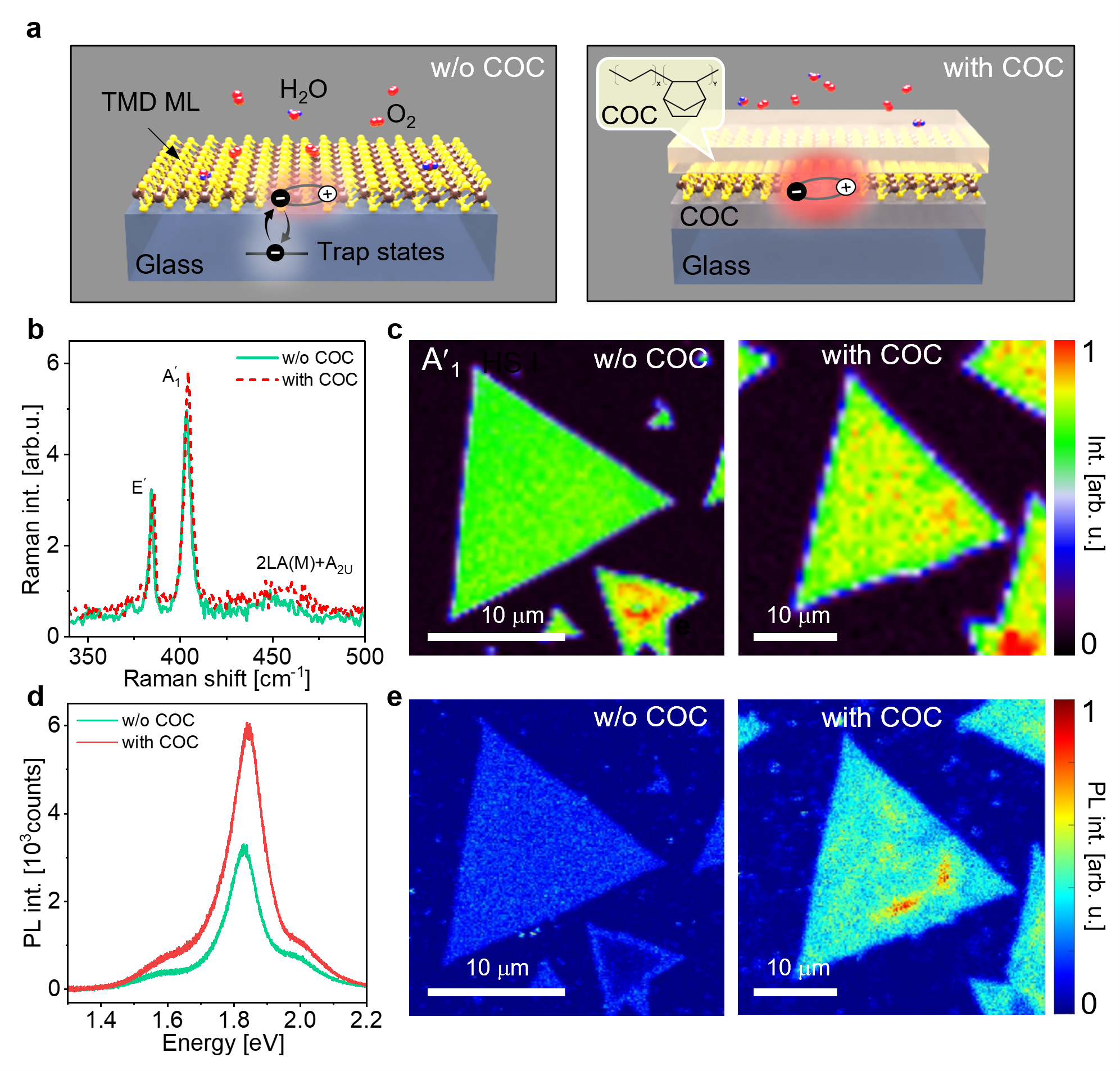}
  \caption{ Schematic, Raman, and PL characterization of CVD-grown $\text{MoS}_2$ ML without and with COC encapsulation. (a) Schematic illustration comparing the structures of the bare $\text{MoS}_2$ on a glass substrate (left panel) and the COC encapsulated $\text{MoS}_2$ ML (right panel). (b) Room-temperature Raman spectra of ML $\text{MoS}_2$ on bare (solid green) and COC-encapsulted substrate (dotted red) (c) Spatial Raman intensity maps of $A'_1$ (\SI{404}{\per\cm}) of the bare $\text{MoS}_2$ (left panel) and the COC-encapsulated $\text{MoS}_2$ (right panel), acquired under 532 nm excitation. (d) Room-temperature PL spectra of ML $\text{MoS}_2$ on bare and COC-encapsulted substrate, recorded under 532 nm excitation (e) PL spatial maps acquired under 532 nm excitation. The encapsulated sample (right panel of e) demonstrates a substantial enhancement in overall PL intensity compared to the bare flake (left panel of e), while also revealing a distinct spatial inhomogeneity across the flake.}
      \label{fig:1}
\end{figure}

\subsection{Cryogenic PL spectroscopy}

To further investigate the excitonic response, PL spectra of both sample configurations were acquired at a temperature of 4 K under {\SI{594}{\nano\meter}} excitation at a power of \SI{50}{\micro\watt}(see Methods  section for details).  Figure 2a shows representative PL spectra for the unecapsulated (top panel) and COC-encapsulated (bottom panel) $\text{MoS}_2$ ML  at 4K. To quantify the individual excitonic contributions, the spectra were fitted using a sum of Lorentzian functions representing the neutral exciton ($X^0$), trion ($X^-$), defect-bound excitons ($X^{L1/L2}$), and, where present, a fine excitonic feature $X^U$ (discussed below). Initial peak positions were guided by literature values, while the final peak positions, linewidths, and amplitudes were allowed to vary within physically reasonable ranges (see Methods section for details). 
At 4 K, COC encapsulation leads to a pronounced increase in the peak PL intensity, exceeding the enhancement observed at room temperature. Averaged over 15 spectra acquired at randomly chosen positions on three spatially separated monolayer crystals, the intensity rises from $(5.4 \pm 1.2)\times10^3$ counts for the bare ML to $(29.8 \pm 3.0)\times10^3$ counts for the COC-encapsulated ML, corresponding to an almost sixfold enhancement. In addition, the neutral exciton exhibits a blue shift of \SI{29.5\pm 0.1}{\milli\electronvolt}  (see green arrow). A narrow additional emission feature at $\sim$\SI{1.96}{\electronvolt}, denoted as $X^U$, consistently appears only in the COC-encapsulated sample. Both samples also show a broad low-energy emission tail, characteristic of localized excitons bound to defects.

\begin{figure}[h]
\centering
  \includegraphics[width=\linewidth]{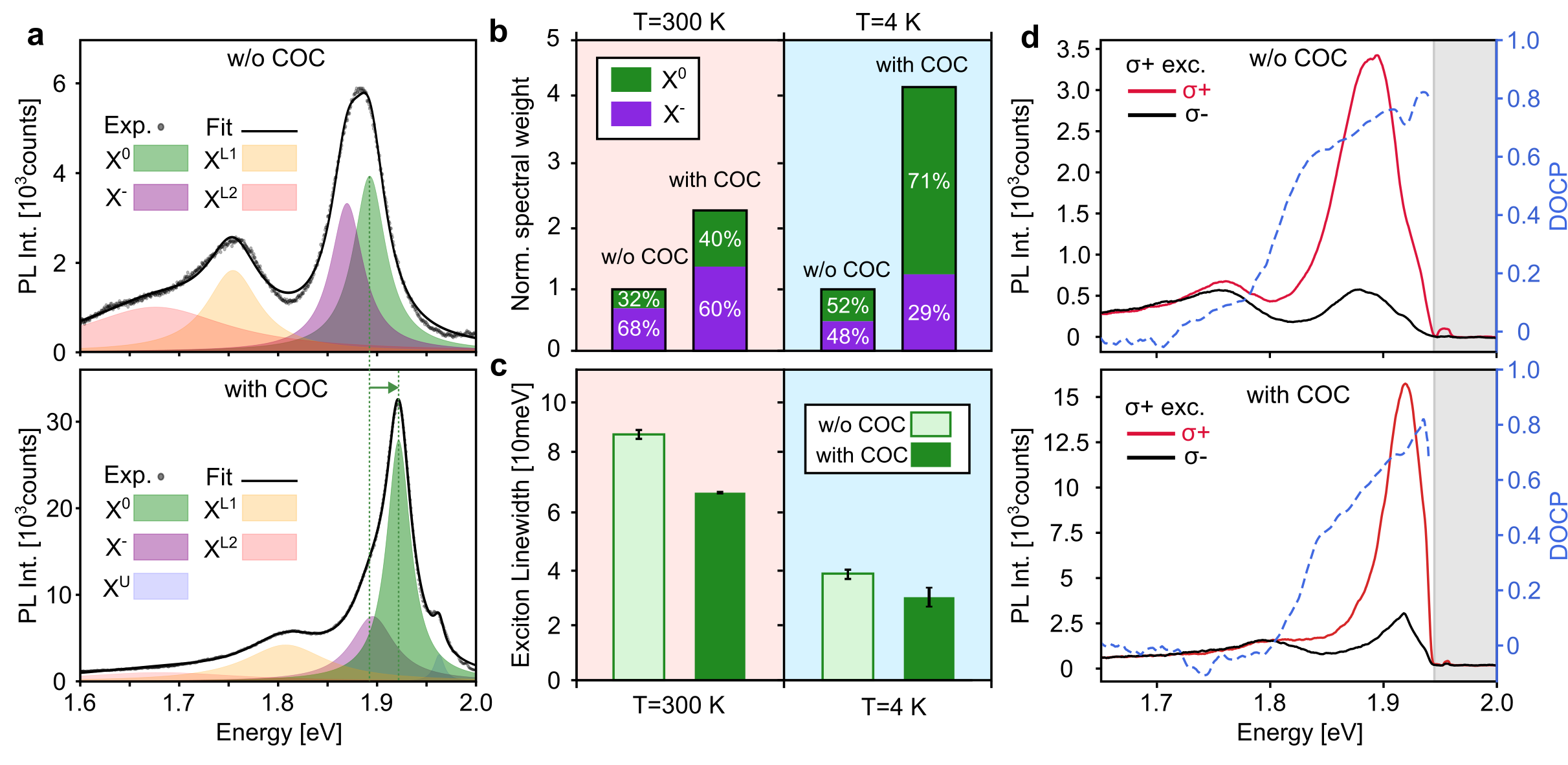}
  \caption{Temperature-dependent optical enhancement and valley coherence of the COC-encapsulated $\text{MoS}_2$ monolayer.
  (a) Low-temperature (\SI{4}{K}) PL spectra comparing the bare and encapsulated samples. Spectral deconvolution reveals the individual contributions of the $X^0$, $X^-$, two low-energy localized defect states $X^L$, and the distinct anomalous high-energy peak, $X^U$. The encapsulated monolayer exhibits a pronounced increase in total PL intensity. (b) Quantitative bar chart summarizing the normalized spectral weight of the individual quasiparticle states at \SI{300}{K} and \SI{4}{K} for both configurations.(c) Comparative bar plot showing the exciton linewidth for the bare and encapsulated samples for both temperatures. The significant reduction of the $X^0$ linewidth for the encapsulated sample indicates suppressed inhomogeneous broadening and a highly uniform dielectric environment. (d) Polarization-resolved PL measurements under resonant excitation ({\SI{633}{\nano\meter}}) at \SI{4}{K} for both the bare $\text{MoS}_2$ on glass and the COC-encapsulated sample. Both configurations demonstrate high optical contrast between the co- and cross-polarized emission channels, confirming robust preservation of valley polarization.}

  \label{fig:2}
\end{figure}
\noindent

Next, we quantify the effect of COC encapsulation on different excitonic complexes by comparing the integrated spectral weights of the neutral exciton ($X^0$) and trion ($X^-$) as obtained from the deconvoluted spectra using numerical integration (Simpson's rule). The resulting relative contributions are summarized in Figure 2b for both room and cryogenic temperatures. At room temperature, the exciton-to-trion ratio changes from 32:68 for the bare ML to 40:60 for the COC-encapsulated sample. The effect becomes even more pronounced at 4 K, where the ratio changes from 52:48 to 71:29. These results indicate that COC encapsulation suppresses trion formation, consistent with reduced substrate-induced doping and a more charge-neutral environment. In addition, the cumulative spectral weight of the excitonic emission increases by approximately a factor of four at 4 K, confirming a substantial enhancement of the radiative recombination efficiency.
The linewidth of the neutral exciton is analyzed in Figure 2c. At 4 K, COC encapsulation reduces the exciton linewidth from approximately 40 meV to 30 meV, indicating suppressed inhomogeneous broadening and a more uniform dielectric environment. A linewidth reduction is also observed at room temperature despite the dominant contribution of exciton-phonon scattering. Together with the increased exciton spectral weight and reduced trion contribution, these results demonstrate that COC effectively suppresses extrinsic disorder and non-radiative decay channels, thereby improving the optical quality of the monolayer. We confirm the reproducibility of the optical properties and the structural integrity of the COC-encapsulated sample by comparing the PL spectra across two independent cryogenic cooling cycles (see Figure S11, SI).

We assess the valley properties by performing polarization resolved PL measurements under resonant (with A-exciton) excitation using
a {\SI{633}{\nano\meter}} CW laser (see Methods section for details). This resonant condition allows for the selective initialization and radiative recombination of valley-polarized excitons. At \SI{4}{K}, the degree of circular polarization defined as $\mathrm{DOCP}=(\mathrm{PL}_{\sigma^+}-\mathrm{PL}_{\sigma^-})/(\mathrm{PL}_{\sigma^+}+\mathrm{PL}_{\sigma^-})$, reaches up to $74\%$ for bare $\text{MoS}_2$ and 72$\%$ for the COC-encapsulated ML when probed at the PL intensity maximum (Figure 2d; filter edge shown in gray).
The high DOCP values observed for both samples indicate efficient preservation of valley polarization. Since the measured DOCP is governed by the competition between exciton decay rate $\gamma_{X}$ and intervalley scattering rate $\gamma_{\mathrm{valley}}$, $\mathrm{DOCP}=1/(1+2\gamma_{\mathrm{valley}}/\gamma_{X})$~\cite{mak2012control}, the persistence of a high DOCP in the COC-encapsulated sample is particularly noteworthy. The reduced excitonic linewidth indicates a lower exciton decay/dephasing rate, which would otherwise be expected to reduce the measured DOCP unless accompanied by a corresponding suppression of intervalley scattering. Thus, the comparable high DOCP suggests that COC encapsulation not only improves the optical quality but also reduces intervalley relaxation processes in the monolayer.

\subsection{Analysis of the A-exciton splitting in the COC encapsulated sample}

\begin{figure}[h]
 \includegraphics[width=1.0\linewidth]{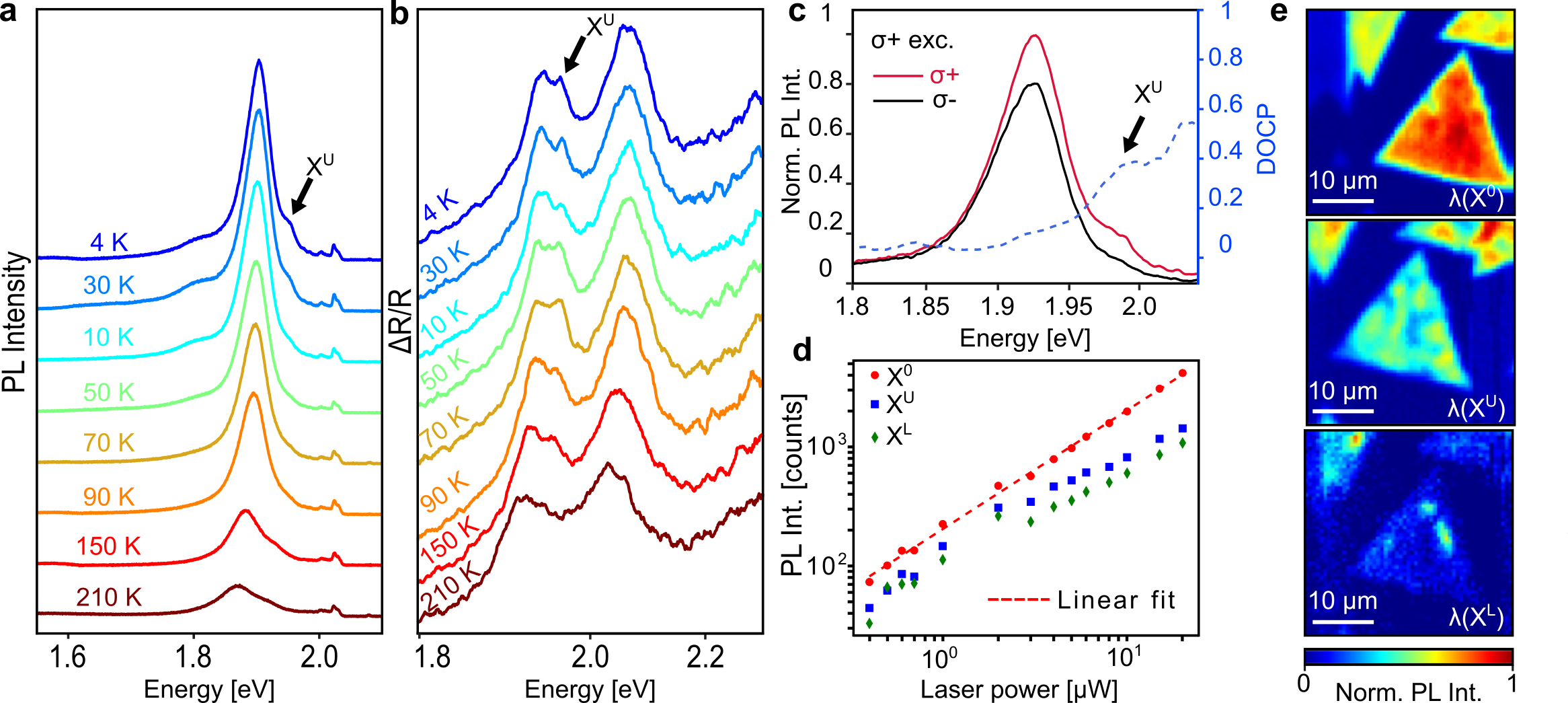}
 \caption{
 Properties of the anomalous excitonic state $X^U$ observed exclusively in the COC-encapsulated sample.(a) Temperature-dependent PL spectra. The black arrow indicates the spectral position  $X^U$ peak. (b) Temperature-dependent differential white-light reflectivity spectra. The black arrow highlight the A-excitons peak splitting. (c) Polarization-resolved PL spectrum obtained under blue-detuned 594 nm excitation highlights at \SI{4}{K}, demonstrating the pronounced degree of circular polarization of the $X^U$ peak. (d) Excitation power dependence of the integrated PL intensities. The $X^0$ emission scales linearly (slope $\approx$ 1), whereas both the $X^U$ peak and the defect-bound exciton $X^{L}$ exhibit  saturation behavior at elevated power. (e) Spatial distribution of distinct excitonic species at \SI{4}{K}. Spectrally filtered PL maps of the encapsulated $\text{MoS}_2$ flake, isolating the integrated intensities of the unknown $\sim$ \SI{1.96}{\electronvolt} transition $\lambda(X^U)$, the neutral exciton $\lambda(X^0)$, and the defect peak $\lambda(X^{L})$.}
  \label{fig:3}
\end{figure}

To elucidate the origin of the additional PL feature at $\sim${\SI{1.96}{\electronvolt} observed exclusively in the COC-encapsulated sample and denoted as $X^U$, we investigated its temperature dependence, valley polarization, power dependence, and spatial distribution.
Figure 3a shows the temperature-dependent PL spectra. With increasing temperature, the $X^U$ peak gradually broadens and merges with the neutral exciton emission, becoming indistinguishable at room temperature (see room-T PL spectrum in Figure 1d). Nevertheless, its continuous spectral evolution suggests that the underlying transition persists over the entire temperature range. A similar behavior is observed in the white-light reflectivity measurements (Figure 3b), which reveal a clear splitting of the A-exciton peak. The splitting remains visible up to approximately 210 K before merging into a single resonance, closely mirroring the PL behavior.

The valley properties of $X^U$ were probed by polarization-resolved PL under blue-detuned \SI{594}{\nano\meter} excitation; resonant excitation at \SI{633}{\nano\meter} could not be used because the emission of interest would overlap with the filter cutoff. As shown in Figure 3c, $X^U$ exhibits a DOCP of up to 40$\%$, indicating partial preservation of valley polarization. In contrast, the neutral exciton shows a low DOCP, as expected under non-resonant excitation~\cite{mak2012control}. Together with its signatures in PL and reflectivity, the finite circular polarization suggests that $X^U$ originates from an excitonic state that retains the spin-valley selection rules of the monolayer.

The excitation-power dependence of the different emission channels ($X^0$, $X^U$, and $X^{L}$) is shown in Fig. 3d. While the neutral exciton $X^0$ exhibits an approximately linear dependence on excitation power ($I \propto P$), both other emissions, $X^{L}$ and $X^U$, show sub-linear behavior and tend to saturate at higher powers. For $X^{L}$, this response is consistent with a finite density of available defect-bound exciton states~\cite{tongay2013defects}. However, sub-linear power dependence is not unique to defect-related emission~\cite{li2019emerging}. More generally, the observed saturation suggests that $X^U$ is not a bright neutral exciton transition, but rather originates from a state with saturation effects, such as a localized excitonic complex or a weakly allowed optical transition.

To assess the spatial character of $X^U$, wavelength-selective PL maps were acquired using narrow bandpass filters that selectively transmit the $X^0$, $X^U$, and defect-related emission band $X^{L}$ (Figure 3e). The corresponding filter transmission curves overlaid with the PL spectrum are shown in Figure S8 of the SI. The resulting maps reveal a spatially inhomogeneous intensity distribution across the crystal for all three emission channels. We note that due to spectral overlap, the $X^U$ map inevitably contains some contribution from the $X^0$ tail. However, their contrasting spatial distributions demonstrate that the $X^U$ map captures a distinct spectral contribution rather than merely reproducing the $X^0$ emission. Unlike the defect-related emission, neither $X^0$ nor $X^U$ exhibits pronounced spatial localization, and both signals remain distributed over the entire monolayer area. In contrast, the defect-related emission $X^{L}$ is concentrated in a few highly localized bright spots, consistent with emission from localized defect states. The similarity between the spatial distributions of $X^0$ and $X^U$ therefore suggests that $X^U$ is unlikely to originate from a strongly localized defect-bound state.

\subsection{Nonlinear characterization}

\begin{figure}[h]
\includegraphics[width=\linewidth]{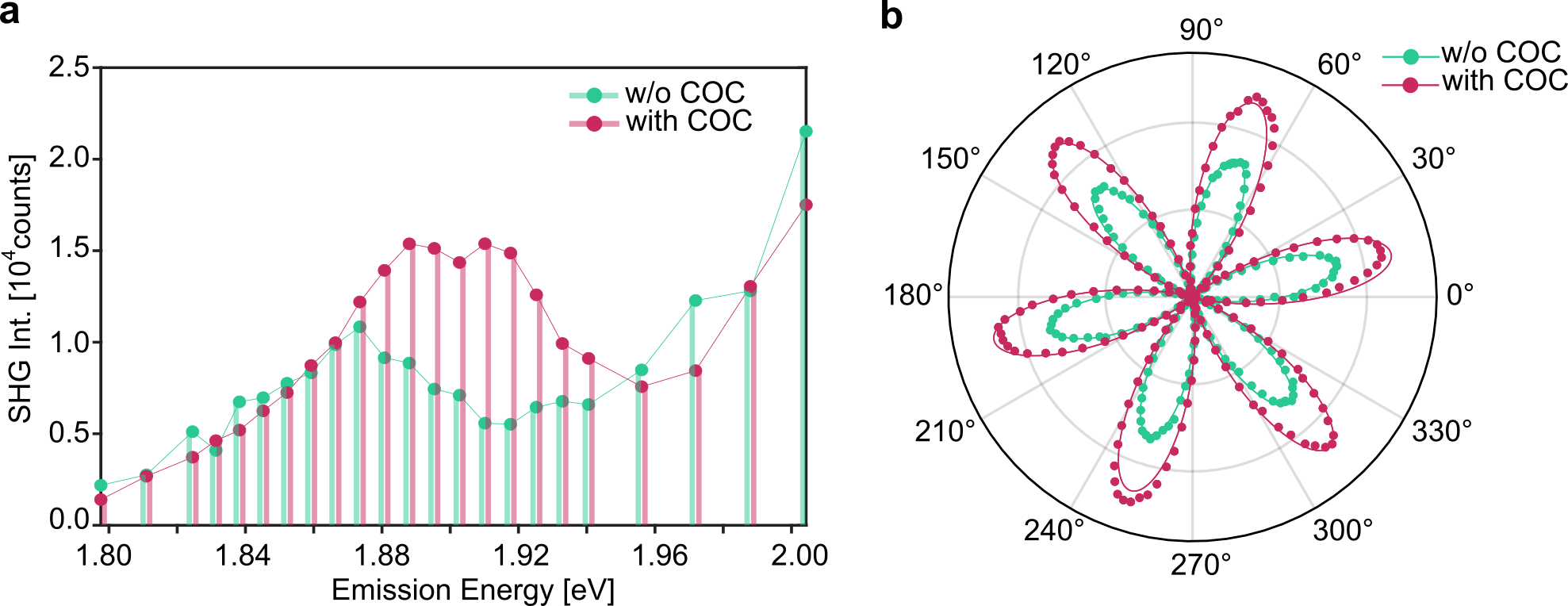}
    %\captionsetup{font = small}
    \caption{Characterization of SHG in bare $\mathrm{Mo}\mathrm{S}_2$ ML on glass substrate (green) and COC-encapsulated $\mathrm{Mo}\mathrm{S}_2$ ML (magenta) at room temperature. (a) SHG intensity as a function of emission energy. (b) Resonant SHG polarization dependence. The excitation energies for COC-encapsulated $\mathrm{Mo}\mathrm{S}_2$ ML and bare $\mathrm{Mo}\mathrm{S}_2$ ML \SI{0.955}{\electronvolt} and \SI{0.937}{\electronvolt}, respectively. Solid lines show the expected SHG polarization dependence of pristine $\mathrm{Mo}\mathrm{S}_2$ ML.}
      \label{fig:4}
\end{figure}

Lastly, we study SHG at room temperature.
Figure \ref{fig:4}a presents SHG excitation spectra of $\mathrm{Mo}\mathrm{S}_2$ ML on glass and COC-encapsulated $\mathrm{Mo}\mathrm{S}_2$ ML on glass, with the fundamental harmonic ranging from $\sim$0.90 to \SI{1.01}{\electronvolt} (1238 to \SI{1378}{\nano\meter}). Compared to the bare $\mathrm{Mo}\mathrm{S}_2$ ML, the COC-encapsulated sample shows an overall intensity increase, a peak blue-shift ($\sim$\SI{40}{\milli\electronvolt}), and a narrower FWHM of the SHG resonance, fully consistent with the PL results (Figure S6, Figure \ref{fig:2}a). These observations indicate improved exciton coherence and enhanced oscillator strength in COC-encapsulated $\mathrm{Mo}\mathrm{S}_2$ MLs \cite{Seyler2015}, arising from the uniform dielectric environment and reduction of charge impurity scattering provided by the encapsulation. 
A resonance at \SI{1.89}{\electronvolt}, approximately \SI{20}{\milli\electronvolt} above the A-exciton, is also observed in the SHG spectrum of the COC-encapsulated $\mathrm{MoS}_2$ ML. Accounting for the room-temperature bandgap renormalization and the corresponding shift of the excitonic resonances, this feature is consistent with the $X^U$ peak observed at low temperature. This result suggests that the underlying excitonic splitting persists up to room temperature. While this splitting is likely obscured by homogeneous and inhomogeneous broadening in room-temperature PL and reflectivity, resonant SHG can exhibit distinct sensitivity to excitonic resonances~\cite{Qian2024}, allowing additional spectral features to be resolved. Spatially-resolved SHG maps acquired at a fundamental photon energy of \SI{0.937}{\electronvolt} (Figure S12, SI) show a largely uniform nonlinear response for both samples. A minor SHG enhancement observed at the lower edge of the COC-encapsulated crystal coincides with the hotspots in the PL map on Figure~\ref{fig:1}e (right) that we attribute to the defect states. 
Figure \ref{fig:4}b shows the characteristic six-fold pattern of SHG polarization-dependent measurement (the corresponding measurement mode is described in e.g. Ref.~\cite{molina2026role}). 
In these measurements, the excitation energies were chosen to match the maxima of the corresponding SHG excitation spectra: \SI{0.955}{\electronvolt} for the COC-encapsulated $\mathrm{MoS}_2$ ML and \SI{0.937}{\electronvolt} for the bare $\mathrm{MoS}_2$ ML.
A moderate deviation between the measured SHG polarization pattern and the theoretical fit is observed for COC-encapsulated $\mathrm{Mo}\mathrm{S}_2$ ML, indicating the presence of strain, which locally breaks the three-fold rotational symmetry of the crystal lattice ($D_{3h}$ point group) \cite{Mennel2018}. The presence of strain is in line with the Raman spectroscopic results.

\subsection{First-Principles Analysis}
\begin{figure}[h]
\centering
\includegraphics[width=1.0\linewidth]{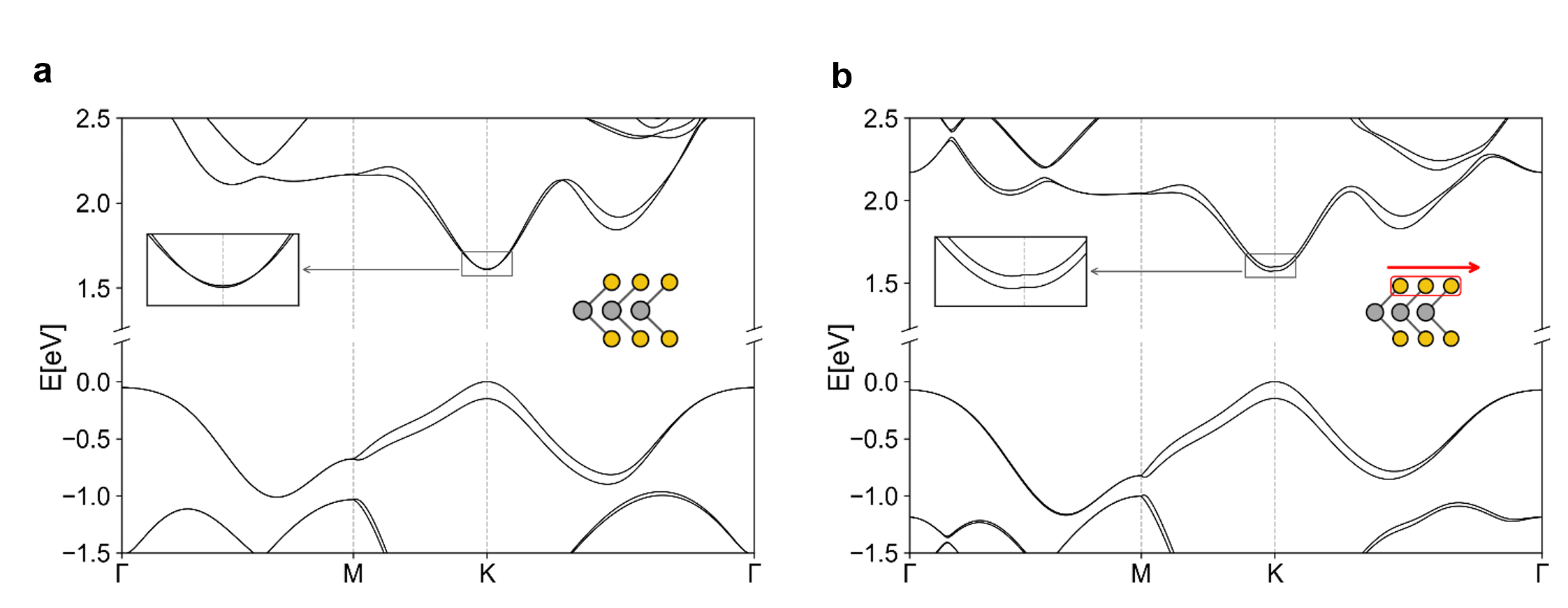}
    %\captionsetup{font = small}
    \caption{DFT band structure of a) pristine $\text{MoS}_2$, b) $\text{MoS}_2$ with the upper S layer horizontally shifted by \SI{0.18}{\text{\AA}} along the direction indicated by the red arrow. The insets magnify the conduction band minima at the K-point revealing the band splitting occurring under the horizonal shift of a single S layer. }
      \label{fig:5}
\end{figure}

To investigate the microscopic origin of the A-exciton splitting, which manifests in the appearance of the $X^U$ exciton state, as well as the overall spectral blueshift of the excitonic features upon COC encapsulation, we investigated the electronic band structure of the system from first principles. Considering that the amorphous polymer interacts directly with the  chalcogen layers, it can introduce local interfacial shear and symmetry-breaking. To model this effect in an exemplary fashion, we simulated a $\text{MoS}_2$ ML with a rigid lateral displacement of a single sulfur layer, contrasting it against its pristine counterpart (Figure 5a-b). As shown in Figure 5b, this asymmetric distortion enhances the splitting of the two lowest states of the conduction band at the K point. Even without displacement, pristine monolayer $\text{MoS}_2$ exhibits a small spin-orbit coupling-induced splitting due to its lack of inversion symmetry (Figure 5a)~\cite{kosmider2013large}. 
 The electronic transitions from the valence-band maximum (VBM) to these lowest unoccupied bands (conduction band minimum CBM and CBM+1) give rise to the A-exciton peak~\cite{ramasubramaniam2012large}. Once the  mirror symmetry ($\sigma_h$) between the two S layers is broken, the corresponding optical transitions become energetically non-degenerate, providing a plausible microscopic origin for the experimentally observed peak splitting.

Motivated by the signatures of compressive strain observed in the Raman spectra and the SHG polarization measurements, we investigated the $\mathrm{Mo}\mathrm{S}_2$ ML under a 0.5$\%$ biaxial compressive strain (Figure S13a), mimicking the global compressive stress exerted by the polymer matrix. This structural modification acts symmetrically on both S sublayers, preserving the near-degeneracy of the lowest unoccupied state at K point, but inducing a rigid upshift of the CBM. By combining both structural effects, namely the horizontal shift of a single S layer and global compressive strain (Figure S14b), our calculations successfully capture the dual mechanisms driving experimental trends. The local asymmetric modification of the S position breaks mirror symmetry and increases the splitting of the fundamental conduction band. the lowest conduction band splitting, yielding the two-peak structure in the A exciton state (see the dependence of the K-valley CBM splitting on chalcogen-layer shift in Figure S15a). Simultaneously, the uniform compressive strain increases the calculated band gap (Figure S15b), accounting for the macroscopically probed spectral blueshift. 

\section{Conclusions}

In conclusion, we demonstrate that COC encapsulation provides a simple, scalable, and low-cost route to substantially improving the optical response of monolayer MoS$_2$ through a straightforward spin-coating process that avoids complex transfer procedures and specialized fabrication infrastructure. COC encapsulation remarkably enhances PL and SHG, reduces the excitonic linewidth, and increases the exciton-to-trion ratio, indicating suppressed non-radiative recombination, reduced substrate-induced doping, and a more homogeneous dielectric environment. At cryogenic temperatures, the high degree of circular polarization observed under resonant excitation demonstrate the preservation of valley-selective optical properties. The improved optical quality further enables the resolution of the A-exciton fine structure and reveals an additional spectral feature (X$^U$), which exhibits signatures in PL, reflectivity, and valley-resolved measurements consistent with a genuine excitonic state associated with the $K/K'$ valleys. Although the microscopic origin of X$^U$ cannot yet be identified unambiguously, its pronounced absorption signature, weak and spectrally narrow emission, finite valley polarization, spatial delocalization, and sub-linear excitation-power dependence are collectively consistent with a weakly allowed excitonic transition, likely arising from a COC-induced modification of the local excitonic landscape.
Our first-principles calculations reveal that a cooperative dual strain profile can drive such spectral changes: a local, asymmetric structural translation at the polymer interface breaks inversion symmetry to lift the conduction band degeneracy while macroscopic uniform compressive strain from the polymer matrix accounts for the concurrent spectral blueshift. Taken together, these results establish COC encapsulation as an effective and scalable platform for enhancing the optical performance of TMD monolayers, uncovering previously unresolved excitonic phenomena, and facilitating their integration into advanced photonic, nonlinear-optical, and valleytronic devices. Moreover, the observed encapsulation-induced spectral modifications suggest that polymer encapsulation may provide a new route toward post-growth engineering of the excitonic and electronic structure of TMDs.

\section{Methods}
\subsection{Sample Preparation}

\threesubsection{CVD growth of $\mathrm{Mo}\mathrm{S}_2$ MLs } 
\justifying
$\mathrm{Mo}\mathrm{S}_2$ MLs were grown on a 300 nm $\mathrm{Si}\mathrm{O}_2$/Si substrate (Siltronix, root mean square (rms) roughness $<$0.2 nm) following a modified CVD growth method reported in Ref.\cite{George2019}.

\threesubsection{Preparation of COC thin film}
\justifying
TOPAS 6013-S04 was used as COC \cite{Kalkan2023}. COC pellets are first washed with toluene and then dissolved in toluene with 0.25 wt $\%$ by stirring for 24 hours. The resulting solution is centrifuged for 10 min at 10000 rpm. Next, they were spin-coated onto the fused silica substrate for 40 seconds at 6000 rpm to get the ultrathin film. The samples were then annealed at 100 $^\circ$C for 3 min for solvent evaporation. A schematic illustration of the COC film preparation process is provided in Figure S1 (SI). The thickness of each COC layer used for the encapsulation is approximately 10 nm. COC thickness as a function of its concentration is shown in Figure S3 (SI).

\threesubsection{Transfer of $\mathrm{Mo}\mathrm{S}_2$ MLs }
\justifying
$\mathrm{Mo}\mathrm{S}_2$ MLs were transferred onto both fused silica and COC-coated fused silica substrates using a wet transfer method\cite{George2019}. A thin layer of poly(methyl methacrylate) (PMMA) was spin-coated onto the $\mathrm{Si}\mathrm{O}_2$/Si substrate with CVD-grown $\mathrm{Mo}\mathrm{S}_2$  MLs. Afterward, the sample (PMMA/$\mathrm{Mo}\mathrm{S}_2$/$\mathrm{Si}\mathrm{O}_2$/Si) was floated on top of a potassium hydroxide (KOH, (85$\%$, Carl Roth)) aqueous solution (2 M) to delaminate the PMMA/$\mathrm{Mo}\mathrm{S}_2$ film from the $\mathrm{Si}\mathrm{O}_2$/Si substrate. The PMMA/$\mathrm{Mo}\mathrm{S}_2$ film was then washed with ultrapure water (18.2 $\text{M}\Omega\cdot\text{cm}$, Membrapure) to remove residual KOH. Afterwards, the PMMA/$\mathrm{Mo}\mathrm{S}_2$ film was placed onto the target substrates and baked at 90 $^\circ$C for 10 min to promote adhesion between the $\mathrm{Mo}\mathrm{S}_2$ and the substrate. Next, the samples were introduced into a $\mathrm{C}\mathrm{O}_2$ critical point dryer (CPD, Tousimis)(a similar description is available in \cite{Rasouli2023}) to dissolve PMMA and to achieve a contamination-free clean surface of the sample before the next COC coating or introducing them into the optical measurements.

\threesubsection{Encapsulation of $\mathrm{Mo}\mathrm{S}_2$  MLs }
\justifying
First, the COC thin film was prepared on a fused silica substrate via spin coating, as described above. $\mathrm{Mo}\mathrm{S}_2$ MLs were subsequently transferred onto the COC-coated substrates using a PMMA-assisted transfer method. To achieve full encapsulation, a second COC layer was spin-coated onto the $\mathrm{Mo}\mathrm{S}_2$/COC/fused silica stack under the same spin-coating and baking conditions used for the initial COC film preparation.

\subsection{\st{} Characterization}
\threesubsection{Cryogenic and Ambient Optical Spectroscopy} 
\justifying
Optical spectroscopy was conducted at both room ($T = 300$ K) and cryogenic temperatures ($T = 4$ K) using an Andor Shamrock 750 spectrometer. All measurements were performed under identical conditions, utilizing a fixed spectrometer slit width of \SI{5}{\micro\meter} and a \SI{5}{\s} exposure time. Low-temperature measurements were performed using a Montana Instruments s50 closed-cycle He cryostat. For steady-state PL measurements shown in Figure 2a and Figures S6, S8-S11 from SI, a LASOS diode-pumped solid-state  laser with a central wavelength of $594$ nm was employed as the excitation source. To isolate the PL signal, the laser line was suppressed using a combination of a $594$ nm notch filter and a $594$ nm long-pass filter (filter's transmission spectrum is shown in Figure S7, SI). 
To account for both sample-to-sample and intra-crystal variations, measurements were performed on three different monolayer crystals located in spatially separated regions of the sample ($>$\SI{200}{\micro\meter} apart), with five positions measured on each crystal. All 15 PL spectra are shown in Figure S9 for the bare $\text{MoS}_2$ and in Figure S10 for the COC encapsulated sample.
Differential reflectivity spectra were acquired using a stabilized tungsten-halogen white light source (see Figure S5).

\threesubsection{Confocal PL Mapping}
\justifying
Spatially resolved PL mapping at room temperature was conducted using a confocal fluorescence microscope (PicoQuant MicroTime 200). A $532$ nm laser was used for excitation and focused onto the sample through a 100$\times$ microscope objective. Surface scans were generated by precisely translating the objective focus across the sample area to obtain high-resolution emission maps.

\threesubsection{Polarization-Resolved Measurements}
\justifying
Polarization-dependent PL spectroscopy was primarily performed using a continuous-wave (CW) He-Ne laser ($\lambda = 633$ nm). To ensure spectral purity, a 633 nm bandpass filter was inserted into the excitation path to eliminate laser side-lobes, while a 632.8 nm long-pass filter was utilized in the detection arm to isolate the PL signal (filter transmission spectrum is shown in Figure S7,SI). Additionally, polarization-resolved measurements were conducted using the 594 nm diode-pumped solid-state laser to specifically investigate the X$^U$ peak observed in the low temperature PL spectra for the encapsulated sample under 594 nm excitation. For all measurements, the incident excitation power on the sample was strictly maintained at \SI{50}{\micro\watt}. To achieve precise circular polarization, the excitation beam was transmitted through a linear polarizer followed by a quarter-wave plate (QWP). The polarization state was verified at the cryostat entry by rotating a secondary linear polarizer from $0^\circ$ to $360^\circ$; the observation of negligible power fluctuations confirmed high-fidelity circular polarization. The Degree of Circular Polarization (DOCP) of the excitation source was further quantified by measuring the maximum ($I_{\text{max}}$) and minimum ($I_{\text{min}}$) intensities using a QWP and linear polarizer combination in the excitation arm, yielding a DOCP of approximately 99 $\%$.
In the detection arm, the emission was analyzed using a QWP followed by a fixed linear polarizer (oriented at $0^\circ$). By switching the QWP between $45^\circ$ and $315^\circ$, the co-polarized ($\sigma^+$) and cross-polarized ($\sigma^-$) components were selectively detected. The signals were imaged onto an Andor spectrometer using a constant slit width of approximately \SI{5}{\micro\meter} to ensure data reproducibility and consistent spectral resolution. All PL spectra were taken with a 5~s exposure time.

\threesubsection{Atomic Force Microscopy}
\justifying
AFM measurements were performed with a Ntegra system (NT-MDT) in tapping mode at ambient conditions using n-doped silicon cantilevers (NSG01, NT-MDT) with a tip radius of 10 nm, with resonant frequencies of 167 kHz.

\threesubsection{Raman and PL Spectroscopy}
\justifying
Raman spectra, room temperature PL spectra as shown in Figure 1d, and Raman mapping were acquired using a Renishaw inVia Raman microscope operated in backscattering geometry with a 532 nm excitation laser (Nd:YAG) under a 100× objective lens (NA = 0.85). 2400 lines/mm grating  with a spacing of $\sim$\SI{0.87}{\per\cm}
between consecutive spectral data points were used for the Raman measurements. For PL spectra, the same 532 nm excitation was used with a grating of 1800 lines/mm and an exposure time of 10 seconds. Automated mapping was enabled by a motorized sample stage, and signals were collected using a CCD detector. The acquired mapping data were analyzed using Renishaw WiRE 5.6 software.

\threesubsection{SHG measurements}
\justifying
All SHG measurements were performed in transmission geometry with emission collected by an electron-multiplying charge-couple device (EMCCD) at room temperature. For SHG spectroscopy, an optical parametric oscillator (Inspire HF100, Spectra-Physics), pumped by a mode-locked Ti:sapphire femtosecond laser (Mai Tai SP, Spectra-Physics) at 820 nm, provided tunable ultrafast fundamental pulses with photon energies ranging from 0.9 to 1.0 eV at a repetition rate of 80 MHz. The pulses were focused to a spot size of approximately 3.7 µm using a 50× near-infrared objective lens (NIR 50×/0.46, Mitutoyo). The SHG signal was collected with a 20× near-infrared objective lens (NIR 20×/0.4, Mitutoyu) and focused on the camera (iXon3 EMCCD, Andor). For polarization-dependent SHG measurements, the fundamental harmonic beam was horizontally polarized by a Glan-Taylor prism (GT10, Thorlabs), whose polarization was then rotated by a motorized half-wave plate (HWP) (AHWP05M-1600, Thorlabs) before entering the focusing objective lens (Figure S13). After collection, the SHG signal and the fundamental harmonic beam passed through a second motorized HWP (2-APW-L2-018B, Altechna), which is simultaneously rotated with the first HWP, and an analyzer (GT10, Thorlabs) oriented perpendicular to the input polarizer. In front of the EMCCD camera, the fundamental harmonic beam was removed by a short-pass filter cutting at 1000 nm (FESH1000, Thorlabs). The six-fold flower patterns were measured at 0.937 eV for the bare $\mathrm{Mo}\mathrm{S}_2$ ML, and at 0.955 eV for the COC-encapsulated $\mathrm{Mo}\mathrm{S}_2$ ML. For both measurements, an average power of 1.45 mW was used before the focusing objective lens.

\subsection{Experimental Data Analysis}
\threesubsection{PL spectra fitting}
\justifying
Starting values for the neutral and charged exciton energies were taken from~\cite{mos2_on_sio2,greben2020intrinsic,mak2013tightly} for the bare $\text{MoS}_2$ on glass. The parameters for the COC encapsulated sample were adapted from hBN-encapsulated systems to account for the modulated dielectric environment ~\cite{uchiyama2019momentum,han2019effects,zhou2025probing,hbn_mos2}.
Room temperature PL spectra for bare $\text{MoS}_2$ sample displayed in Figure S6 were fitted using pseudo-Voigt functions~\cite{bender2024spectroscopic}. The trion peak ($X^-$) was initialized at $1.820\text{ eV}$ and the neutral exciton ($X^0$) at $1.850\text{ eV}$, and B exciton ($X^B$) was initialized at $2.020\text{ eV}$ and for COC encapsulated sample the trion peak ($X^-$) was initialized at $1.850\text{ eV}$ and the neutral exciton ($X^0$) at $1.885\text{ eV}$, and B exciton ($X^B$) was initialized at $2.020\text{ eV}$.

Low temperature PL spectra for bare $\text{MoS}_2$ sample were fitted to a 4-Lorentzian model. The main emission profiles were constrained using initial peak positions for cryogenic $\text{MoS}_2$ emission: the trion ($X^-$) was initialized at $1.860\text{ eV}$ (bounds: $1.845–1.875\text{ eV}$, and the neutral exciton ($X^0$) at $1.895\text{ eV}$ (bounds: $1.885–1.910\text{ eV}$. Two additional broad Lorentzian sub-peaks were initialized at lower energies ($1.760\text{ eV}$ and $1.680\text{ eV}$) to accommodate defect-bound or localized tail states. Following convergence, the integrated areas of the isolated $X^0$ and $X^-$ Lorentzian components were calculated via Simpson's numerical integration to evaluate their relative population statistics.
For the COC encapsulated sample, five distinct electronic transitions characteristic of encapsulated cryogenic $\text{MoS}_2$ monolayer samples were defined with tailored initialization parameters: a high-energy shoulder(the unknown peak $X^U$)  at $1.960\text{ eV}$ (bounds: $1.940–1.980\text{ eV}$), the neutral exciton ($X^0$) at $1.920\text{ eV}$ (bounds: $1.910–1.940\text{ eV}$), and the trion ($X^-$) at $1.890\text{ eV}$ (bounds: $1.870–1.910\text{ eV}$). Two broader sub-peaks located at lower energies ($1.810\text{ eV}$ and $1.720\text{ eV}$) were incorporated to fit defect-bound and localized states.

\subsection{DFT}
DFT calculations were performed using the Quantum ESPRESSO suite~\cite{giannozzi2017advanced}. Relaxation calculations for the strained configurations used a 9x9x1 $k$-mesh, while the SCF calculations were performed using a 36x36x1 $k$-mesh and included spin-orbit coupling through a fully-relativistic, norm-conserving pseudopotential ~\cite{schlipf2015optimization}. A vacuum layer of 20 {\text{\AA}} was included in the out-of-plane direction to effectively decouple periodic replicas. The self-consistency energy threshold was set to $10^{-8}$ {\text{Ry}}, while interatomic forces were minimized below $10^{-3}$ $\text{Ry}/\text{bohr}$. The generalized-gradient approximation (Perdew-Burke-Ernzerhof functional~\cite{perdew1996generalized}) was adopted for the exchange correlation potential. While semi-local DFT is well-known to underestimate absolute electronic band gaps, it reliably captures structural relaxation, strain potentials, and orbitally driven band-splitting trends in TMDs~\cite{crowley2016resolution, mori2008localization, borlido2020exchange}. Because the focus of this theoretical analysis is to identify the qualitative symmetry-breaking and strain mechanisms driving experimental spectral features, rather than predicting absolute quasiparticle energies, the adopted computational settings provide an efficient and physically robust framework for this work.

\medskip
\textbf{Acknowledgements} \par %delete if not applicable))

The authors acknowledge financial support from the Deutsche Forschungsgemeinschaft (DFG, German Research Foundation) through the Collaborative Research Centre CRC/SFB 1375 NOA - Nonlinear Optics down to Atomic Scales (Project B2, and A8 Project No. 398816777), the International Research Training Group IRTG 2675 Meta-Active (Project No. 437527638), and Research Training Group RTG 3014 PhInt - Photo-Polarizable Interfaces and Membranes (Project No. 521747072). T.H., A.P., B.A., C.C., and A.T. acknowledge financial support from the DFG Priority Programme SPP 2244 2DMP (Project TU149/21-1, Project No. 535253440). T.H., A.P., B.A., and A.T. acknowledge financial support from the European Innovation Council (EIC) through the project 2DSPIN-TECH (No. 101135853). C.C. acknowledges additional funding from the DFG (Project No.547611111).
S.D. is part of the Max Planck School of Photonics supported by the Dieter Schwarz Foundation, the German Federal Ministry of Research, Technology and Space (BMFTR), and the Max Planck Society.

% References
\medskip

\bibliographystyle{MSP}
\bibliography{references}

\end{document}